\documentstyle[11pt,epsfig]{article}
\newcommand{\bildchen}[3]{%                           % 
\begin{center}                                        %
\begin{flushleft}                                     %
\makebox{\Large $\displaystyle #2$}                   %
\end{flushleft}                                       %
\mbox{{\epsfig{figure=#1,width=12.cm,%                %
bbllx=1.8cm,bblly=9.2cm,bburx=20.cm,bbury=19.cm}}}    %
\end{center}                                          %
\begin{flushright}                                    %
   {\Large $\displaystyle #3$ \hspace*{5ex}}          %                  
\end{flushright}}                                     %
\begin{document}
\begin{titlepage}
\begin{center}
April, 1996      \hfill     HUTP-96/A015 \\
\vskip 0.2 in
{\large \bf A NEW LATTICE ACTION FOR STUDYING TOPOLOGICAL CHARGE}\footnotetext{~}

\vskip .2 in
       {\bf  Pilar Hern\'andez}\footnote{e-mail: hernandez@huhepl.harvard.edu.}\\
and \\
       {\bf   Raman Sundrum}\footnote{e-mail: sundrum@huhepl.harvard.edu.}
        \vskip 0.3 cm
       {\it Lyman Laboratory of Physics \\
Harvard University \\
Cambridge, MA 02138, USA}
 \vskip 0.7 cm 
%\vskip 0.3 in

\begin{abstract}
We propose a new lattice action for non-abelian gauge theories, which will 
reduce short-range lattice artifacts in the computation of the topological
susceptibility. The standard Wilson action is replaced by the Wilson action 
of a gauge covariant interpolation of the original fields to a finer 
lattice. If the latter is fine enough, the action of
all configurations with non-zero topological charge will satisfy the 
continuum bound. 
As a simpler example we consider the $O(3)$ $\sigma$-model in two dimensions, 
where a numerical analysis of discretized continuum instantons indicates that
a finer lattice with half the lattice spacing of the original is enough to 
satisfy the continuum bound.    
\end{abstract}
\end{center}

\end{titlepage}

Field configurations with non-zero topological charge are expected to have a strong influence on the dynamics of asymptotically free theories. In QCD, such 
configurations are
responsible for breaking axial symmetry and resolving the U(1) problem 
\cite{oldies}. The study of these effects however requires non-perturbative techniques and one would
expect that ultimately Monte Carlo methods on the lattice would be best suited to it. The observable to consider is inspired by the  
classic large-$N_c$ analyses of Witten and of Veneziano \cite{Nc}, which showed 
that, 
\begin{eqnarray}
m^2_{\eta'} + m^2_{\eta} - 2\;m^2_K = \frac{6 \;\chi_t}{f^2_\pi},   
\label{eta}
\end{eqnarray}
where $\chi_t$ is the topological susceptibility. In the continuum it is
given by,
\begin{eqnarray}
\chi_t \equiv \int d^4 x < q(x) q(0) > |_{no\; quarks}
\label{chi}
\end{eqnarray} 
with $q(x)$ being the topological charge density,
\begin{eqnarray}
q(x) =  \frac{1}{32 \pi^2} \epsilon_{\mu\nu\rho\sigma} Tr[\;F_{\mu\nu}\;  F_{\rho\sigma}].
\label{contq}
\end{eqnarray}
The topological charge, $Q \equiv \int q(x)$, is an integer if the field 
strength vanishes at infinity or if (euclidean) space-time is compact.   
A continuum analysis also shows that the action of any configuration with 
non-zero topological charge must satisfy the following bound, 
\begin{eqnarray}
S \geq \frac{8 \pi^2 |Q|}{g_0^2}.
\label{contbound}
\end{eqnarray}

A big effort has been devoted to the study of the topological susceptibility 
on the lattice. There are several choices for the operator $q(x)$. The naive
discretization of (\ref{contq}) does not yield integer values for $Q$ and 
requires renormalization factors \cite{venelat}. On the other hand, the cleaner geometrical definition due to L\"uscher \cite{luscher3} gives an integer-valued topological 
charge and does not require renormalization. Here we will deal only with a geometrical definition very similar to 
L\"uscher's. The geometrical topological susceptibility is then obtained by computing,
\begin{eqnarray}
\chi_t = \frac{< Q^2 >}{V},
\end{eqnarray}   
where $V$ is the volume of the lattice. This definition is clearly equivalent 
to (\ref{chi}) in finite volume.  
 From the continuum formula (\ref{eta}), it follows that 
the topological susceptibility in QCD should scale as $(mass)^4$ in the 
continuum limit, namely
\begin{eqnarray}
\chi_t \sim (\;b^{-1}  exp^{-\frac{1}{2 \beta_0 g_0^2}} \;)^4,
\end{eqnarray}
where $b$ is the lattice spacing and $\beta(g_0)\simeq -\beta_0 g_0^3$ is the leading term of the beta function. 

 However, it was found, first in the $O(3)$ $\sigma$-model in two 
space-time dimensions \cite{luscher1}\cite{luscher2}
and then in non-abelian gauge theories in four \cite{gksw1} \cite{teper}, that
 the standard actions in both cases give rise to short-range fluctuations with 
non-zero geometrical topological charge and smaller action than the continuum bound. 
These fluctuations, often referred to as ``dislocations'', overwhelm the
contribution of the slowly varying fields, which would otherwise dominate in
the continuum limit, and are expected to destroy the scaling of the topological
susceptibility. On the other hand, if the bound is satisfied, the semiclassical continuum analysis of non-abelian gauge theories
indicates that the susceptibility should indeed be ultraviolet finite, and therefore scale. Satisfying the continuum bound is thus a sufficient condition
for scaling in non-abelian gauge theories \cite{luscher2}\cite{gksw1}\cite{teper}\cite{improved}
\footnote{It has been argued by L\"uscher\cite{luscher2} that in the case 
of the $O(3)$ $\sigma$-model, the failure in finding scaling is not related to 
the existence of dislocations and is an
essential problem of this model: the partition functional in 
the instanton sector shows a logarithmic ultraviolet divergence in the continuum analysis.}. This suggests that eq. (\ref{contbound}) can be satisfied on the lattice by giving dislocations a larger action.

Several proposals to solve this problem have been considered in the literature, 
like the cooling method \cite{teper} and the use of improved actions \cite{gksw1} \cite{improved}. In this paper we propose to use a new action which satisfies the continuum bound. It is 
related in spirit to the actions proposed in \cite{improved}, but may be 
simpler to implement. 

The idea behind our new action is easy to understand
once one realizes the mismatch between the geometrical definition of 
topological 
charge \cite{luscher3} and the Wilson action for gauge fields which 
allows dislocations to arise.  Consider a continuum instanton $A_\mu(y)$ which saturates the 
bound (\ref{contbound}), and discretize it 
on a lattice of spacing $b$,
\begin{eqnarray}
U_\mu(s) \equiv P \exp( i\; \int^{s b+ b \hat{\mu}}_{s b} dy \;A_\mu(y)\;),
\label{dis}
\end{eqnarray}
where $s\; b$ are the sites of the $b$ lattice.
The geometrical definition of topological charge 
assigns a non-zero value 
even to a lattice configuration (\ref{dis}) obtained from very small instantons, of order of the lattice spacing $b$. On the other hand, it is clear that the Wilson action very poorly approximates the continuum 
action of the original instanton of $O(b)$, and 
in fact it is generically smaller and therefore does not satisfy the bound.
Such rough configurations are dislocations and can destroy the scaling of the susceptibility. 
An important observation is that the geometrical 
topological charge assigned to a lattice configuration can be understood as  the naive topological 
charge of a {\it continuum} configuration obtained by smoothly interpolating 
the lattice configuration. Then it is clear that if, instead
of using the standard Wilson action of the original lattice configuration, we 
use the continuum action of the interpolated configuration, the continuum bound 
is necessarily satisfied, as first suggested in \cite{gksw}.

More concretely, in \cite{us2} we described a procedure to obtain a continuum gauge field
$a_{\mu}(y)$ which interpolates any $b$-lattice configuration\footnote{The continuum field is
 differentiable inside each $b$-hypercube and transversely 
continous across the $b$-boundaries, and when discretized according 
to (\ref{dis}) gives back $U_\mu(s)$ \cite{us2}.}. The interpolation 
is local and gauge covariant, i.e. for a $b$-lattice gauge transformation $\Omega(s)$,
\begin{eqnarray}
a_\mu[U^\Omega] = a_\mu^\omega[U],
\label{gi}
\end{eqnarray} 
where $\omega$ is a gauge transformation in the continuum.  (Other interpolations can 
also be used in this context \cite{gksw}.) 
A  geometrical
topological charge of the $b$-lattice configuration is defined as the one 
associated to the interpolated field \cite{us2},
\begin{eqnarray}
Q =  \frac{1}{32 \pi^2} \int d^4 y \; Tr[\; \tilde{f}_{\mu\nu}(y) f_{\mu\nu}(y) ]. 
\end{eqnarray}
 This definition has the same properties as L\"uscher's original definition \cite{luscher3} and requires roughly the same computational effort.
Now, it is clear that replacing the standard Wilson action by the continuum action of $a_\mu$, i.e.
\begin{eqnarray} 
S^{cont} = \frac{1}{2 g_0^2} \int d^4 y \;Tr[ f_{\mu\nu}(y) f_{\mu\nu}(y)],
\end{eqnarray}
insures that the continuum bound is satisfied for the same
reason as it is in the continuum. Notice that 
this is a perfectly gauge invariant action for the lattice field $U_\mu(s)$, by
eq. (\ref{gi}). 

However, for numerical purposes the continuum action is impractical \cite{gksw}. The central observation of this paper is that it is not necessary to interpolate all the 
way to the continuum, but just to a finer lattice, with lattice 
spacing $f$ (we will take $b/f$ to be integer). From now on, we refer to $x$ as the sites on the $f$-lattice and $s$ as the sites on the 
$b$-lattice ( $x_\mu = s_\mu + m_\mu \frac{f}{b}$, $m_\mu = 0,..., \frac{b}{f}-1$). The interpolation procedure 
in \cite{us2} gives a set of link variables $u_\mu[U](x)$, such that
\begin{eqnarray} 
- \frac{i}{f}\; log(u_\mu(x)) = a_\mu(x) + O(f/b^2),
\label{imp}
\end{eqnarray}
 where $a_\mu(x)$ is the 
continuum interpolation discussed above, at point $x$. 
On the $f$ lattice, we can simply choose the standard Wilson action. The partition functional will then have the form, 
\begin{eqnarray}
{\cal Z} =  \int \prod_s {\cal D} U(s) \;e^{- S^f_{wilson}[u[U]]}, 
\label{effact}
\end{eqnarray}
where, $D U$ is the usual Haar measure for non-abelian gauge fields on the $b$-lattice, and the Wilson action in terms of the interpolated link variables $u[U]$ is given by,  
\begin{eqnarray}
S^f_{wilson}[u[U]] = \frac{1}{g_0^2} \sum_x \sum_{\mu\neq\nu} (I - u_{\mu\nu}[x] + h.c. ),
\nonumber\\
u_{\mu\nu}[x] \equiv Tr[u_\mu(x) 
u_\nu(x+\hat{\mu})
u_\mu^\dagger(x+\hat{\nu}) u_\nu^\dagger(x)].
\label{actna}
\end{eqnarray}
Again, this action is gauge invariant, because the functional $u[U]$ is gauge 
covariant \cite{us2}. From eq. (\ref{imp}) it then follows that,
\begin{eqnarray}
S^f_{wilson} = \frac{1}{2 g_0^2} \int d^4 y \;Tr[ f_{\mu\nu}(y) f_{\mu\nu}(y)] + O(f/b).
\end{eqnarray}
Although the $O(f/b)$ terms are not necessarily positive definite, clearly by 
taking $f/b$ small enough we can come arbitrarily close to satisfying the continuum bound, so that $\chi_t$ shows scaling.
Determining how small this ratio must be in practice requires a numerical 
analysis, which is beyond the 
scope of this letter. However, we have carried out a numerical analysis 
of this issue in a simplified model in two dimensions, which shares many 
features with four dimensional Yang-Mills theories and our results encourage
us to believe that $\frac{f}{b}$ need not be very small in order
to recover scaling of $\chi_t$.  

\section{O(3) $\sigma$-Model in 2D}
 
 The O(3) $\sigma$-model in two dimensions \cite{o3ins} \cite{o3} is the simplest asymptotically free field theory and, as is well-known, it has instanton solutions 
\cite{o3ins}. A continuum bound on the action exists for configurations with non-zero topological charge. We will show that, while in the standard lattice formulation of this model \cite{luscher1}, there are topologically non-trivial 
configurations with a smaller action than the continuum bound, this is not the
case when we use an improved action along the lines
described above. For a similar discussion in the context of renormalization 
improved actions see \cite{o3imp}. We will not address in this paper the interesting question of whether short-range configurations satisfying the continuum bound in this model can still dominate in the continuum limit, as argued by L\"uscher \cite{luscher2}. 
 
The $O(3)$ $\sigma$-model in the continuum is defined by the action,
\begin{eqnarray}
S= \frac{1}{2 g_0^2} \int d^2 x \sum_\mu \; (\partial_\mu \vec{N} (x))^2,
\end{eqnarray}
where $\vec{N}$ is a 3-component real field satisfying the constraint ${\vec{N}\;} ^2 = 1$ (which defines the surface of a sphere of unit radius). The continuum topological charge in this model is given by,
\begin{eqnarray}
Q = \frac{1}{8 \pi} \int d^2 x \; \epsilon_{\mu\nu} \vec{N} \cdot
 ( \partial_\mu \vec{N} \times \partial_\nu \vec{N}),
\end{eqnarray}
which is the number of times that space-time wraps around the $\vec{N}$-sphere.
Using the identity \cite{poly},
\begin{eqnarray}
\frac{1}{4}  (\partial_\mu \vec{N} + \epsilon_{\mu\nu} \vec{N} \times \partial_\nu \vec{N})^2 = \frac{1}{2} (\partial_\mu \vec{N})^2 
-\frac{1}{2} \;\epsilon_{\mu\nu} \vec{N} \cdot \partial_\mu \vec{N} \times
\partial_\nu \vec{N},
\label{trick}
\end{eqnarray} 
it follows that, 
\begin{eqnarray}
S = \frac{4 \pi Q}{g_0^2} + \frac{1}{4 g_0^2} \int d^2 x \;( \partial_\mu \vec{N}
+ \epsilon_{\mu\nu} [ \vec{N} \times \partial_\nu \vec{N}])^2, 
\end{eqnarray}
and, since the second term is positive definite,
\begin{eqnarray}
S \geq \frac{4 \pi |Q|}{g_0^2}.
\label{o3bound}
\end{eqnarray}
Continuum instantons saturate this bound \cite{o3ins}.
 
In a standard lattice treatment the action is,
\begin{eqnarray}
S^b = \frac{1}{2 g_0^2} \sum_s \sum_\mu (\hat{\partial}_\mu \vec{N}(s))^2
\label{o3stan}
\end{eqnarray}
where $s$ are the sites of a two dimensional lattice and  
$\hat{\partial}_\mu \vec{N} \equiv \vec{N}(s +\hat{\mu}) - \vec{N}(s)$.
The topological charge as defined by Berg and L\"uscher \cite{luscher2} is given by, 
\begin{eqnarray}
Q^b = \sum_s q(s),
\end{eqnarray}
\begin{eqnarray}
q(s) = \frac{1}{2\pi} \{Im[Log[f[s,s+\hat{\mu},s+\hat{\mu}+\hat{\nu}] \;
f[s,s+\hat{\mu}+\hat{\nu},s+\hat{\nu}]]] \} ,\nonumber\\
f[s_1,s_2,s_3] \equiv \frac{ 1+ \sum_{i<j} \vec{N}(s_i)\cdot \vec{N}(s_j) + i \vec{N}(s_1) \cdot[ \vec{N}(s_2) \times \vec{N}(s_3) ]}{[2 \; \prod_{i<j} (1+ \vec{N}(s_i)\cdot \vec{N}(s_j))]^{1/2}} \nonumber\\
\label{topo3}
\end{eqnarray}
It is not hard to understand this formula.
Consider the plaquette $(s, \hat{1},\hat{2})$. 
The spin variables at the corners are 
four points on the sphere. We can form two triangles with corners at these points and with sides along geodesics,  
$T_1(s) = (\vec{N}(s),\vec{N}(s+\hat{1}),\vec{N}(s+\hat{1}+\hat{2}))$ and  $T_2(s) = (\vec{N}(s),
\vec{N}(s+\hat{1}+\hat{2}),\vec{N}(s+\hat{2}))$. $q(s)$ is simply the sum of the area of these two image triangles on the $\vec{N}$-sphere.  
If periodic boundary conditions are imposed,  
$Q^b$  is necessarily an integer. 

 With these definitions of the action (\ref{o3stan}) and the topological charge (\ref{topo3}), the analysis in \cite{luscher1}\cite{luscher2} showed that there are 
dislocations, i.e. configurations with non-zero topological charge that 
have a smaller action than the continuum bound (\ref{o3bound}). We will show that 
this picture changes considerably when the action of an interpolation 
of $\vec{N}$ is used.  

We first interpolate the 
lattice spin variables to a finer lattice, by moving along geodesics on the
sphere.  The interpolation will be done locally, i.e. the interpolated fields
within a plaquette $(s, \hat{1}, \hat{2})$ of the $b$ lattice will only depend on the four spins
associated with this plaquette, i.e. $\vec{N}(s), \vec{N}(s+\hat{1}), \vec{N}(s+\hat{2}), \vec{N}(s+\hat{1}+\hat{2})$. 
The points of the finer lattice contained in the plaquette at $s$, are $x = s + t_1 \hat{1} + t_2 \hat{2}$, with $0 \leq t_{1,2} \leq 1$ and multiples of $f/b$. The spin fields at these points will be
denoted by $\vec{n}(t_1, t_2)$. We first define the interpolated spin fields
at the corners in the obvious way,
\begin{eqnarray}
\vec{n}(t_1, t_2) = \vec{N}(s + t_1 \hat{1} + t_2 \hat{2}) \;\;\;\;\;\;\;\;  t_1, t_2 = 0, 1.
\end{eqnarray}
We now interpolate along the one dimensional boundaries of the plaquette, 
by moving along geodesics on the sphere. For $0 < t_1 < 1$,   
\begin{eqnarray}
\vec{n}(t_1, 0) = \frac{\sin[\theta(0,0; 1,0) ( 1- t_1)]}{\sin[\theta(0,0; 1,0)]} \vec{n}(0, 0) + \frac{\sin[\theta(0,0; 1,0)\; t_1]}{\sin[\theta(0,0; 1,0)]} \vec{n}(1, 0)\nonumber
\end{eqnarray}
\begin{eqnarray}
\vec{n}(t_1, 1) = \frac{\sin[\theta(0,1;1,1) ( 1- t_1)]}{\sin[\theta(0,1;1,1)]} \vec{n}(0, 1) + \frac{\sin[\theta(0,1;1,1)\; t_1]}{\sin[\theta(0,1;1,1)]} \vec{n}(1, 1),\nonumber\\
\end{eqnarray}
where we have defined, 
\begin{eqnarray} 
\theta(t_1,t_2; t'_1,t'_2) \equiv d\; [ \vec{n}(t_1, t_2), \vec{n}(t'_1, t'_2)] = \arccos[\vec{n}(t_1, t_2)\cdot \vec{n}(t'_1, t'_2)].
\end{eqnarray}
The expression for spins on the boundaries along direction $\hat{2}$ are analogous so we skip the formulae. 

We now define the spins in the interior. In order to do this we first 
interpolate along the diagonal geodesic connecting $\vec{N}(s)$ and $\vec{N}(s +\hat{1} +\hat{2})$,
\begin{eqnarray}
\vec{n}(t, t) = \frac{\sin[\theta(0,0; 1,1) ( 1- t)]}{\sin[\theta(0,0; 1,1)]} \vec{n}(0, 0) + \frac{\sin[\theta(0,0; 1,1) t]}{\sin[\theta(0,0; 1,1)]} \vec{n}(1, 1).
\end{eqnarray}
For the interior points with $t_1 \geq t_2$, we obtain the spin by simply 
moving a fraction $t_2$ of the distance along the geodesic from $\vec{n}(t_1, 0)$ to $\vec{n}(t_1, t_1)$.
Similarly for points with $t_2 \geq t_1$, the spin variable is obtained 
by moving a distance $t_2$ along the geodesic linking
$\vec{n}(t_1, 1)$ and $\vec{n}(t_1,  t_1)$. This corresponds to 
separately interpolating the interior points of the two image triangles $T_1(s)$ and $T_2(s)$. 
Defining,
\begin{eqnarray}
\alpha(t_1) \equiv \theta(t_1,0; t_1, t_1) \;\;\;\; \beta(t_1) \equiv \theta(t_1, t_1; t_1, 1),
\end{eqnarray}
the final expression is,
\begin{eqnarray}
\vec{n}(t_1, t_2) = \frac{\sin[ \alpha\; (1- t_2/t_1)]}{\sin[\alpha]} \vec{n}(t_1, 0) + \frac{\sin[\alpha\; t_2/t_1]}{\sin[\alpha]}\vec{n}(t_1, t_1)  \;\;\; t_1 \geq t_2 \nonumber\\
\vec{n}(t_1, t_2) = \frac{\sin[ \beta\; (1- \tilde{t}_2/\tilde{t}_1) ]}{\sin[\beta]} \vec{n}(t_1, 1) + \frac{\sin[\beta\; \tilde{t}_2/\tilde{t}_1]}{\sin[\beta]} 
\vec{n}(t_1, t_1)  \;\;\; t_2 \geq t_1
\label{interpol}
\end{eqnarray}
with $\tilde{t}_i \equiv 1 - t_i$.

The interpolation (\ref{interpol}) is now a configuration on an $f$-lattice and we define the improved action to be,
\begin{eqnarray}
S^{f} = \frac{1}{2 g_0^2} \sum_x \sum_{\mu =1,2}  (\hat{\partial}_\mu \vec{n}(x))^2,
\end{eqnarray}
where $x$ is any point on the $f$-lattice and $\hat{\partial}_\mu \vec{n}(x) \equiv \vec{n}(x+ f/b \hat{\mu}) - \vec{n}(x) = O(f/b)$.

Now, we can easily derive a bound on the action, noticing that by construction, 
\begin{eqnarray}
Q^f = Q^b,
\end{eqnarray} 
where $Q^f$ is the topological charge for the $f$-lattice configuration: it
is given by (\ref{topo3}), simply substituting $\vec{N}$ by $\vec{n}$ and 
$s$ by $x$. Now we can expand $Q^f$ in $\hat{\partial} \;\vec{n} \sim O(f)$ and find that,
\begin{eqnarray}
Q^f = \frac{1}{8 \pi} \sum_x \epsilon_{\mu\nu} \;\vec{n}(x) \cdot (\hat{\partial}_\mu \vec{n}(x) \times \hat{\partial}_\nu \vec{n}(x)) + O(f/b).
\end{eqnarray}
Using (\ref{trick}) one finds,
\begin{eqnarray}
S^{f} = \frac{4\pi Q^f}{g_0^2} + \frac{1}{4 g_0^2} \sum_x \;( \hat{\partial}_\mu \vec{n}
+ \epsilon_{\mu\nu} [ \vec{n} \times \hat{\partial}_\nu \vec{n}])^2 + O(f/b).
\end{eqnarray}
Then,    
\begin{eqnarray}
S^{f} \geq \frac{4 \pi Q^b}{g_0^2} + O(f/b).
\label{latbound}  
\end{eqnarray}
Since it is not possible to prove that the $O(f/b)$ terms are positive definite, we do not know how small an $f/b$ we need to be sufficiently close to 
the continuum bound. 

In order to address this question, we considered the discretization of a continuum instanton configuration with 
unit topological charge given by \cite{poly},
\begin{eqnarray}
n_1 + i n_2 = 2 w/(1+|w|^2) \;\;\;\; n_3 = (1- |w|^2)/(1+|w|^2)\nonumber
\end{eqnarray}
\begin{eqnarray}
w(x_1 + i x_2) \equiv \frac{x_1 + i x_2 - a (1 + i)}{x_1 + i x_2 - c (1+ i)}
\end{eqnarray} 
where $\vec{n} = (n_1, n_2, n_3)$. We take $a$ and $c$ to be real and 
define, 
\begin{eqnarray}
a = (r_0 +r/2 \sqrt{2}) \;\;\;\;\;\; c = (r_0 - r/2 \sqrt{2})
\end{eqnarray} 
The radius of the instanton is proportional to $r = |a - c|$, while the center is located 
at $(r_0, r_0)$, with $r_0 = |a + c|/2$. Generically there is always 
a critical, $r_c$, below which 
$Q^b = 0$. We expect, and find numerically, that $r_c \sim b$. We consider the center to be situated at the center of the volume to reduce finite 
volume effects. Obviously, the continuum instanton is not 
periodic. We impose periodic boundary conditions by defining, 
\begin{eqnarray}
\vec{N}(s_1,s_2) \equiv \vec{n}(z_1, z_2), \;\;\;\;\;z_i = \frac{2 L_{max}}{\pi} \frac{\sin(\pi(1 - s_i/L_{max}))}{1-\cos(\pi(1 - s_i/L_{max}))},
\label{ste}
\end{eqnarray}
where $z_i$ are coordinates on the infinite plane and $s_i \in (-L_{max}, L_{max})$ are the coordinates
on the lattice (torus). $(2 L_{max})^2$ is
then the number of lattice sites. The connection between these two sets of coordinates 
is established through sterographic projection.  This deformation of the infinite volume instanton
is small if the instanton size is much smaller than $L_{max}$. For such 
configurations therefore, the continuum action nearly saturates the bound. 
An alternative procedure would be
to discretize the continuum instanton solutions of this model in a torus 
\cite{o3imp}. 
 
Figure 1 sumarizes our results. It represents the action of the discretized instanton configuration as the radius is varied. The continuous line corresponds to 
the standard action (\ref{o3stan}), while the dashed lines correspond
to the improved action for different values of the ratio $f/b$. It is
clear that the standard action is problematic, since for $r > r_c$ the action
is smaller than the continuum bound.  For the new 
action however the situation is different. Not only is the continuum
bound satisfied, but also as the instanton is shrunk to sizes of $O(b)$, a small barrier develops, separating the 
$Q^b =0$ and $Q^b = 1$ sectors. The existence of this barrier is easy 
to understand. The new action is the action of   
the instanton which has been first discretized on the $b$-lattice and then 
interpolated. 
For a large instanton (compared to $b$), the interpolation should recover approximately the original 
configuration and so the action must be near that of a continuum instanton and
show scale invariance. On the other hand, 
for an instanton of size $\sim b$ a lot of information is lost in the discretization and 
the interpolation is not expected to give a configuration 
similar to the original continuum instanton. In general, the interpolation of 
small discretized instantons of $O(b)$ will then be some other configuration that need not be
even approximately an instanton, and consequently its action will be larger than the 
bound, since only instantons saturate the bound. Thus the action must increase
as we decrease $r$ near $r_c$, as is clearly seen in Fig. 1.   
The other important point to notice is that the continuum bound (\ref{o3bound}) is satisfied even for a ratio $f/b$ as large as $1/2$. This indicates that the extra effort required 
to use the improved action is clearly managable in this case.

\begin{figure} 
%\begin{center}
%\mbox{
%\epsfig{file=o3_topo_50_05.eps,height=15cm}}
%\end{center}
\bildchen{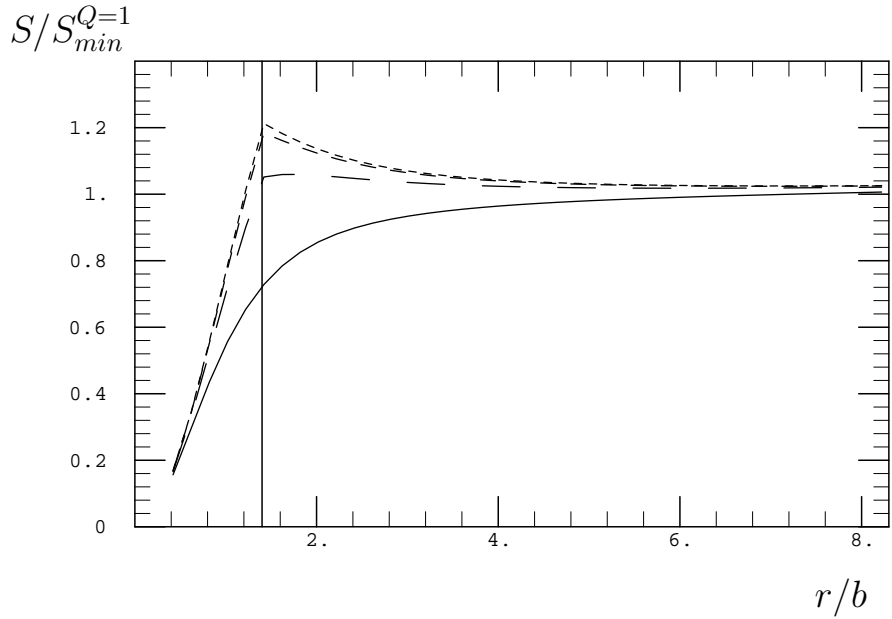}{S/S^{Q=1}_{min}}{r/b}
 \caption[]{Action of a discretized instanton of the O(3) model (normalized
to the continuum one-instanton bound (\ref{o3bound})) as
a function of its radius, $r$. The full line is the standard action in a $100 \times 100$ lattice and
the dashed lines correspond respectively to $f/b =1/2, 1/4, 1/6$
(smaller $f/b$, smaller dashing). The vertical line at $r_c \sim 1.4 b$ 
separates the $Q^b = 0, 1$ sectors.}
\label{fig:lattice}
\end{figure}

\section{Conclusions}

We have presented a new action for non-abelian gauge theories which 
is better suited to studying topology on the lattice than the standard
Wilson action. The idea is to use the Wilson action 
of an interpolation to a finer lattice of the original 
lattice configuration. In this way, if the ratio 
of lattice spacings is small enough, 
the continuum bound on configurations with non-zero topological charge is satisfied. 
In non-abelian gauge theories this is a sufficient condition for proper
scaling of the topological susceptibility. We also considered a new 
action for the $O(3)$ $\sigma$-model in two dimensions along the same lines. 
A numerical study of discretized continuum instantons in this model indicates that 
already for a ratio of lattice spacings of $1/2$, the continuum bound is satisfied. Although the results for the $O(3)$ model are very promising, a separate 
numerical analysis is needed in the Yang-Mills case to determine the ratio 
there.

\section*{Acknowledgments}

This research was supported by NSF-PHYS-92-18167. We thank U.J. Wiese
for discussions. P.H. is supported by the Harvard Society of Fellows and 
also partially by the Milton Fund of Harvard University.

\end{document}